\documentclass[iop]{emulateapj}
\usepackage{graphicx}
\usepackage{epstopdf}

\tighten

\newcommand{\asec}{\hbox to 1pt{}\rlap{$^{\prime\prime}$}.\hbox to 2pt{}}
\newcommand{\amin}{\hbox to 1pt{}\rlap{$^{\prime}$}.\hbox to 2pt{}}

\slugcomment{Accepted for publication in the {\it The Astrophysical Journal}, 24 July 2012}
\shortauthors{Postman et al.}
\shorttitle{The Core of A2261-BCG}

\begin{document}

\title{A Brightest Cluster Galaxy with an Extremely Large Flat Core}

\author{Marc Postman\altaffilmark{1},
Tod R. Lauer\altaffilmark{2},
Megan Donahue\altaffilmark{3},
Genevieve Graves\altaffilmark{4},
Dan Coe\altaffilmark{1},
John Moustakas\altaffilmark{5,6},
Anton Koekemoer\altaffilmark{1},
Larry Bradley\altaffilmark{1},
Holland C. Ford\altaffilmark{7},
Claudio Grillo\altaffilmark{8},
Adi Zitrin\altaffilmark{9},
Doron Lemze\altaffilmark{7},
Tom Broadhurst\altaffilmark{10,11},
Leonidas Moustakas\altaffilmark{12},
Bego\~na Ascaso\altaffilmark{13},
Elinor Medezinski\altaffilmark{7},
\& Daniel Kelson\altaffilmark{14}
}

\altaffiltext{1}{Space Telescope Science Institute, 3700 San Martin Drive,
Baltimore, MD 21208, USA}
\altaffiltext{2}{National Optical Astronomy Observatory, P.O. Box 26732,
Tucson, AZ 85726, USA}
\altaffiltext{3}{Dept. of Physics and Astronomy,  Michigan State University,
East Lansing, MI 48824, USA}
\altaffiltext{4}{Dept. of Astronomy, 601 Campbell Hall,
University of California, Berkeley, CA 94720, USA}
\altaffiltext{5}{Center for Astrophysics and Space Sciences,
University of California, La Jolla, CA 92093, USA}
\altaffiltext{6}{Dept. of Physics and Astronomy, Siena College,
515 Loudon Road, Loudonville, NY 12211, USA}
\altaffiltext{7}{Dept. of Physics \& Astronomy, Johns Hopkins University,
3400 N. Charles Street, Baltimore, MD  21218, USA}
\altaffiltext{8}{Dark Cosmology Centre, Niels Bohr Institute, University of Copenhagen, 
Juliane Mariesvej 30, 2100 Copenhagen, Denmark}
\altaffiltext{9}{University of Heidelberg, Albert-Ueberle-Str. 2,
69120 Heidelberg, Germany}
\altaffiltext{10}{Dept. of Theoretical Physics,
Univ. of the Basque Country UPV/EHU, Bizkaia, 48940 Leioa, Spain}
\altaffiltext{11}{IKERBASQUE, Basque Foundation for Science,
Alameda Urquijo 36-5, 48008 Bilbao, Spain}
\altaffiltext{12}{Jet Propulsion Laboratory,
California Institute of Technology, 4800 Oak Grove Dr., Pasadena, CA 91109, USA}
\altaffiltext{13}{Instituto de Astrof\'isica de
Andaluc\'ia (CSIC), C/Camino Bajo de Hu\'etor, 24, Granada 18008, Spain}
\altaffiltext{14}{The Observatories of the Carnegie Institution of Washington, 
813 Santa Barbara Street, Pasadena, CA 91101, USA}

\begin{abstract}

Hubble Space Telescope images of the galaxy cluster Abell 2261, obtained as part
of the Cluster Lensing And Supernova survey with Hubble, 
show that the brightest galaxy in the cluster, A2261-BCG,
has the largest core yet detected in any galaxy.  The cusp radius
of A2261-BCG is 3.2 kpc, twice as big as the next largest core known,
and $\sim3\times$ bigger than those typically seen in the most luminous BCGs.
The morphology of the core in A2261-BCG is also unusual, having a
completely flat interior surface brightness profile,
rather than the typical shallow cusp rising into the center.  This implies
that the galaxy has a core with constant or even centrally decreasing
stellar density.  Interpretation of the core as an end product of the
``scouring'' action of a binary supermassive black hole implies
a total black hole mass $\sim10^{10}M_\odot$ from
the extrapolation of most relationships between core structure and
black hole mass.
The core falls $1\sigma$ above the cusp-radius versus
galaxy luminosity relation.
Its large size in real terms, and the extremely large black hole mass required
to generate it, raise the possibility that the core has been
enlarged by additional processes, such as the ejection of the black
holes that originally generated the core.  The flat central
stellar density profile is consistent with this hypothesis.  The
core is also displaced by 0.7 kpc from the center of the surrounding
envelope, consistent with a local dynamical perturbation of the core.

\end{abstract}


\section{A Large Core As a Test of Core Formation} 

Brightest cluster galaxies and most luminous early-type galaxies
brighter than $M_V \sim-21$ have ``cores'' in
their central starlight distributions \citep{f97, laine, l07a}.
Cores are marked by a distinct physical radius
interior to which the projected starlight
surface-brightness increases only slowly as $r\rightarrow0,$
in marked contrast to the surrounding envelope, which has
a much steeper profile (in logarithmic units).
In qualitative terms, a core looks like a central ``plateau''
in the starlight distribution.
In more quantitative terms,
a core can be defined as the central region of a galaxy
where the surface brightness takes the form of a shallow cusp,
$I(r)\propto r^{-\gamma},$ with $\gamma<0.3$
as $r\rightarrow0$ \citep{l95,l05}.
Importantly, galaxies fainter than  $M_V \sim-21$ generally do not have
cores, having $\gamma>0.5$ instead, as $r\rightarrow0.$
This distinction is of physical interest, as the
presence or absence of a core correlates with
the strength of the stellar rotation field, isophote shape,
nuclear radio emission, and overall X-ray emission,
in addition to the total galaxy luminosity \citep{f97,l07b}.

The formation of cores has long been thought to be due to the action
of black holes on the central structure of galaxies.
Their form and size may reflect both the mass of the central
black hole in the galaxies, and the merger history that created the galaxies.
\citet{bbr} hypothesized that a binary black hole
created in the merger of two galaxies would eject stars from the center of
the newly created system as the binary slowly hardened.
In simple terms, the black hole binary ``scours'' out the
center of the galaxy, thus ``flattening'' the otherwise steeply
rising central starlight distribution as $r\rightarrow0.$
Subsequent N-body simulations have demonstrated this
phenomenon, directly \citep{ebi, mak97, mnm}.

\citet{f97} showed that cores occur in the most luminous elliptical galaxies,
and are correlated with slow-rotation and ``boxy'' isophotes in these galaxies,
concluding that core formation is a natural end-point of
dissipationless mergers of two progenitor elliptical galaxies.
The conclusion that nearly every elliptical galaxy
has a black hole at its center \citep{mag},
coupled with the conclusion that
the most massive elliptical galaxies were formed
by merging pre-existing gas-free galaxies, explains why cores are
found in nearly all luminous ellipticals.

While ``core scouring'' is an attractive hypothesis
for the formation of cores, there may be additional mechanisms for
binary black holes to generate cores.  \citet{rr} suggested that when the
two black holes in the binary ultimately merge, asymmetric emission of
gravitational radiation could eject the merged hole from the
center of the galaxy, causing the center to ``rebound'' in response
to the large reduction in central mass. \citet{dm04}, \citet{bk04},
and \citet{gm08} studied this problem in detail for realistic galaxy
models, demonstrating that the ejection of the merged black hole indeed
could cause the central distribution of starlight to re-adjust such that
it would create a core in the projected stellar surface brightness profile.
An interesting ancillary effect discussed in these works is the possibility
that the ejected black hole would remain bound to the host galaxy on
a radial orbit. In that case, it would repeatedly fall through the
center of the galaxy, continuing to enlarge the core through
dynamical friction.

Some observational support for the scouring origin of cores
comes from the measurement of core ``mass deficits'' compared
to the black hole masses in the same galaxies \citep{f97}.
The mass deficit, $M_d,$ is the inferred
mass of stars ejected from the center of the
galaxy requited to create a core,
and is estimated by a reference to a postulated initial
form of the galaxy fitted to the envelope, such as a S\' ersic law
\citep{gr04}.
Various studies estimating mass deficits \citep{f97,milo,rav02,gr04,m06,l07a,kb}
typically find $M_d\propto M_\bullet,$ the black hole mass, with the
constant of proportionality of order unity, but with large scatter
about the mean relation.

The observational context for understanding the large core in the brightest cluster galaxy in Abell 2261 
(hereafter referred to as A2261-BCG)
is provided by the extensive work done with
the {\it Hubble Space Telescope} ({\it HST})
on the structure of nearby galaxies.
\citet{l07a} constructed a large sample of galaxies that had
high-resolution surface photometry obtained with {\it HST;}
significantly, it included the core parameters measured by
\citet{laine}, who studied a large sample of brightest cluster galaxies.
Very few galaxies have cores as large as 1 kpc; the largest core in the
\citet{l07a} sample is that for NGC 6166 = A2199-BCG,
which has a core size of $\sim1.5$ kpc.
\citet{mac} drew attention to the large core that they discovered in BCG
MS0735.6+7421; however, it is compatible with the largest BCG cores
measured in the \citet{laine} sample.
The core in A2261-BCG, however, is over twice as large as that in NGC 6166.
It provides an extreme test of the possible
mechanisms hypothesized to generate cores.

\begin{figure*}
\plotone{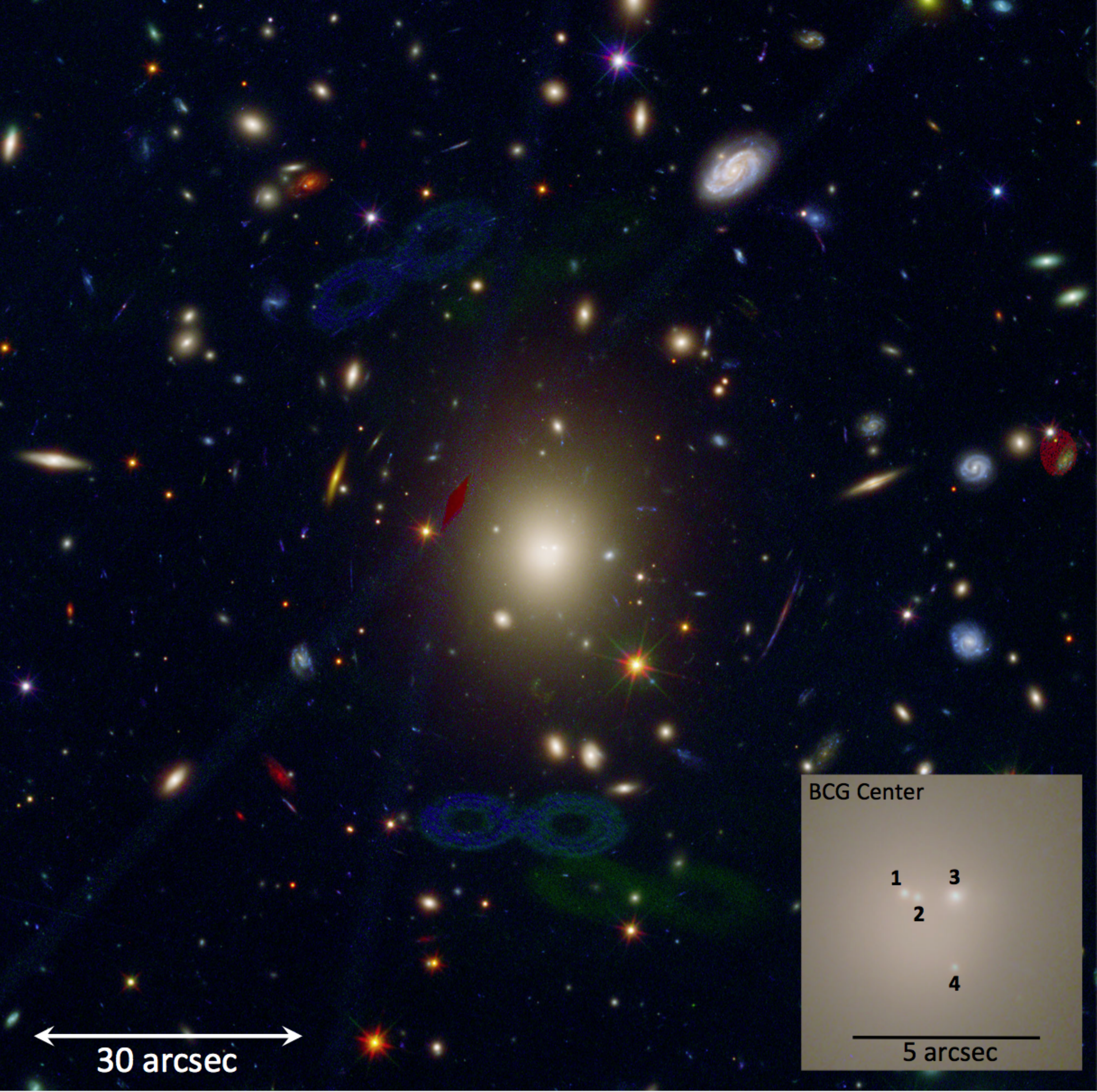}
\caption{Color composite HST image, from CLASH ACS/WFC and WFC3/IR images, showing the BCG in A2261 and its neighbors 
in the central $2 \times 2$ arcminute region of the cluster. The insert
in the lower right hand corner shows a zoomed in region centered on the 
BCG with contrast adjusted to highlight the bright knots (labelled 1,2,3,4) in the core. The
orientation is north up and west to the right. The faint ``figure 8" patterns at the 6 o'clock and 11 o'clock positions are due to internal reflections in the ACS camera of light from a nearby bright star. The red ``diamond" at the 10 o'clock position near the BCG is caused by 
a gap in areal coverage due to the multiple orientations used in the CLASH survey. The red ``blob" at the right edge of the image is a WFC3/IR detector artifact that does not easily calibrate out.}
\label{fig:a2261}
\end{figure*}

\section{Observations and Properties of A2261-BCG}

The center of Abell 2261 was observed for a total of 20 orbits as part
of the Cluster Lensing And Supernova survey with Hubble (CLASH)
multi-cycle treasury program between
March 9, 2011 and May 21, 2011 in 16 broadband filters 
from $0.22~-~1.6~\mu{\rm m}$ \citep{clash}.  
For the calculation of physical quantities, we 
assume a cosmology with $\Omega_m = 0.3,\ \Omega_{\Lambda} = 0.7,$
and $H_0 = 70$ km s$^{-1}$ Mpc$^{-1}$.
At the mean redshift of Abell 2261, $z = 0.2248$ \citep{Coe12},
$1''$ subtends 3.61 kpc and the distance modulus is 40.241.


Abell 2261 is included in the CLASH X-ray selected subsample of 20 clusters.
The CLASH X-ray selected sample consists of clusters with X-ray
temperatures greater than 5 keV and exhibit a high degree of dynamical
relaxation as evidenced by {\it Chandra X-ray Observatory}
images that show well-defined central surface brightness peaks
and nearly concentric isophotes.
The intracluster medium (ICM) of Abell 2261, in particular, is characterized
by an X-ray temperature of T$_x = 7.6\pm 0.30$ keV, a bolometric X-ray 
luminosity of $1.80\pm 0.20 \times 10^{45}$ erg s$^{-1}$ and a [Fe/H] 
abundance ratio that is $0.31\pm 0.06$ times the solar value \citep{ag89}. 
The estimated mass within r$_{2500}$ (radius where the density is
2500 times the critical density) is $M_{2500} = (2.9\pm 0.5)\times 10^{14}
~\rm{M}_\odot$ with a gas fraction of $0.115\pm 0.01$.

\citet{M08} found Abell 2261 to have a small level of substructure
in its X-ray gas surface brightness distribution and \citet{G09}
classified the cluster as ``disturbed." \citet{Coe12} find 
that application of the caustic technique
(e.g., \citealt{Diaferio97,Diaferio05}) to spectroscopically 
measured galaxies in the vicinity of Abell 2261 suggests that the
dynamical center of the cluster is located $\sim6'$ ($\sim$1.3 Mpc) south of the BCG.
However, the BCG ($z = 0.2233$) in Abell 2261
is at equatorial coordinates of 17:22:27.18 $+$32:07:57.1 (J2000), 
putting it within $1\asec6$ 
(5.8 kpc) of the centroid of the ICM X-ray emission.
Its mean velocity offset relative to that of the cluster mean redshift
is 367 km s$^{-1}$. The distribution of BCG velocity offsets for a sample of 42 Abell clusters
has a mean value of 264 km s$^{-1}$ \citep{pl}. An expanded dataset of 174 Abell clusters, each
with at least 50 spectroscopically confirmed members, yields a mean value of the BCG velocity
offset of 234 km s$^{-1}$, with 22\% having offsets of at least 350 km s$^{-1}$.
The A2261-BCG, thus, appears to be reasonably aligned with
the center of the cluster's main gravitational potential well.

The unusual core of A2261-BCG was discovered
during the initial inspection of the CLASH {\it HST} images of A2261.
Visually, the core presents itself as a large, round, uniform disk of
low surface brightness, as can be seen in
Figure~\ref{fig:a2261}, which shows a color composite of the central
$2'\times 2'$ region of Abell 2261 made from the CLASH images.
In addition, four compact sources with colors similar to
the BCG itself are superimposed on the outskirts of the core.
The 3 brightest knots are marginally detected ($<3\sigma$) in the 
WFC3/UVIS F336W image, well detected ($>6\sigma$) at longer wavelengths,
and not detected in either the F225W or F275W UVIS passbands.

\subsection{Nuclear Activity}

Understanding the central structure of A2261-BCG
requires knowledge of whether or not the core
hosts a central supermassive black hole.
This problem will be considered at length later in the paper,
but we note here that the center of the galaxy does harbor a
radio source that is at least an order of magnitude more
powerful than those associated with star forming regions,
and is thus evidence in favor of an active galactic nucleus (AGN).
The NRAO VLA Sky Survey \citep[NVSS;][]{nvss} finds a source
at 1.4 GHz with an integrated flux of $5.3 \pm 0.5$ mJy
that lies $6\asec3$ east of the BCG position.
No other NVSS sources are found within $2\amin45$ of the BCG.
Data from the Faint Images of the Radio Sky at Twenty-Centimeters
(FIRST) survey \citep{first} reveals a 3.4 mJy source that is
more coincident with the core, lying
$1\asec6$ west of the BCG position.
The FIRST detection limit at the source position is 0.99 mJy per beam. No other 
FIRST sources are found within $1\amin92$ of the BCG.
Presumably the NVSS and FIRST detections closest to the BCG correspond to
the same source. The two radio positions are $7\asec5$ apart,
which is less than a $2\sigma$ difference and is
consistent with the different astrometric uncertainties for the two surveys.
Unfortunately, the VLA sky survey data currently do not have sufficient
angular resolution to determine whether one of the knots,
or the core itself, may host the AGN (VLA FIRST resolution is $5''$, NVSS resolution is $45''$).

If the radio source is at the cluster redshift then the FIRST detection yields
an absolute luminosity of $7.8\times10^{39} {\rm~erg~s^{-1}}$ at 1.4GHz 
corresponding to $L_{1.4GHz} = 5\times10^{23}~{\rm W~ Hz^{-1}}$.
\citet{hl12} also find a central source
with luminosity $5.0\times10^{39} {\rm~erg~s^{-1}}$ at 5 GHz
corresponding to $L_{5.0GHz} = 1\times10^{23}~{\rm W~ Hz^{-1}}$.
The radio fluxes are indicative of a mildly powerful FR I radio source
typical of those seen in BCGs \citep[e.g.][]{croft}

Apart from the radio source,
there is also a faint $24~\mu{\rm m}$  detection of 0.44 mJy
\citep{hoff}, which would indicate either star formation
or a hot dust torus around an AGN.  As in the case of the radio
observations, however, the angular resolution is probably too poor
to distinguish whether or not the flux comes from the center
of the core or any of the associated knots.

In contrast, optical spectra do not reveal any significant
evidence of emission lines that may be associated with an AGN
or cooling flow activity (the spectra are presented in the next subsection).
Further, there are no known X-ray or EUV sources identified with any of
the compact knots; the X-ray limit on any point source in the core
is $L_{2-10{\rm keV}} < 3.8\times10^{41}~{\rm erg~s^{-1}}$ \citep{hl12}.
At the same time, it is typical for central radio source in BCGs
not to have strong X-ray counterparts.
A black hole mass of $\sim 10^{10}~M_\odot$ is in general consistent with
the power requirements associated with the ICM shock fronts
seen in other clusters \citep[e.g.][]{mac}; but for such high mass
black holes, the power needed for such shocks
is only $\sim 0.001~L/L_{ed},$ which therefore implies that the black holes
can be radiatively inefficient. \citet{hl12} note that synchrotron cooling can 
strongly affect the X-ray luminosity from the most massive black holes since the
cooling break occurs below X-ray wavelengths, making such massive black holes
underluminous at X-ray wavelengths compared to their radio luminosity. 
For A2261-BCG in particular, \citet{hl12} estimate an
Eddingtion ratio of $\sim10^{-6}~-~10^{-8}.$

\begin{figure}
\plotone{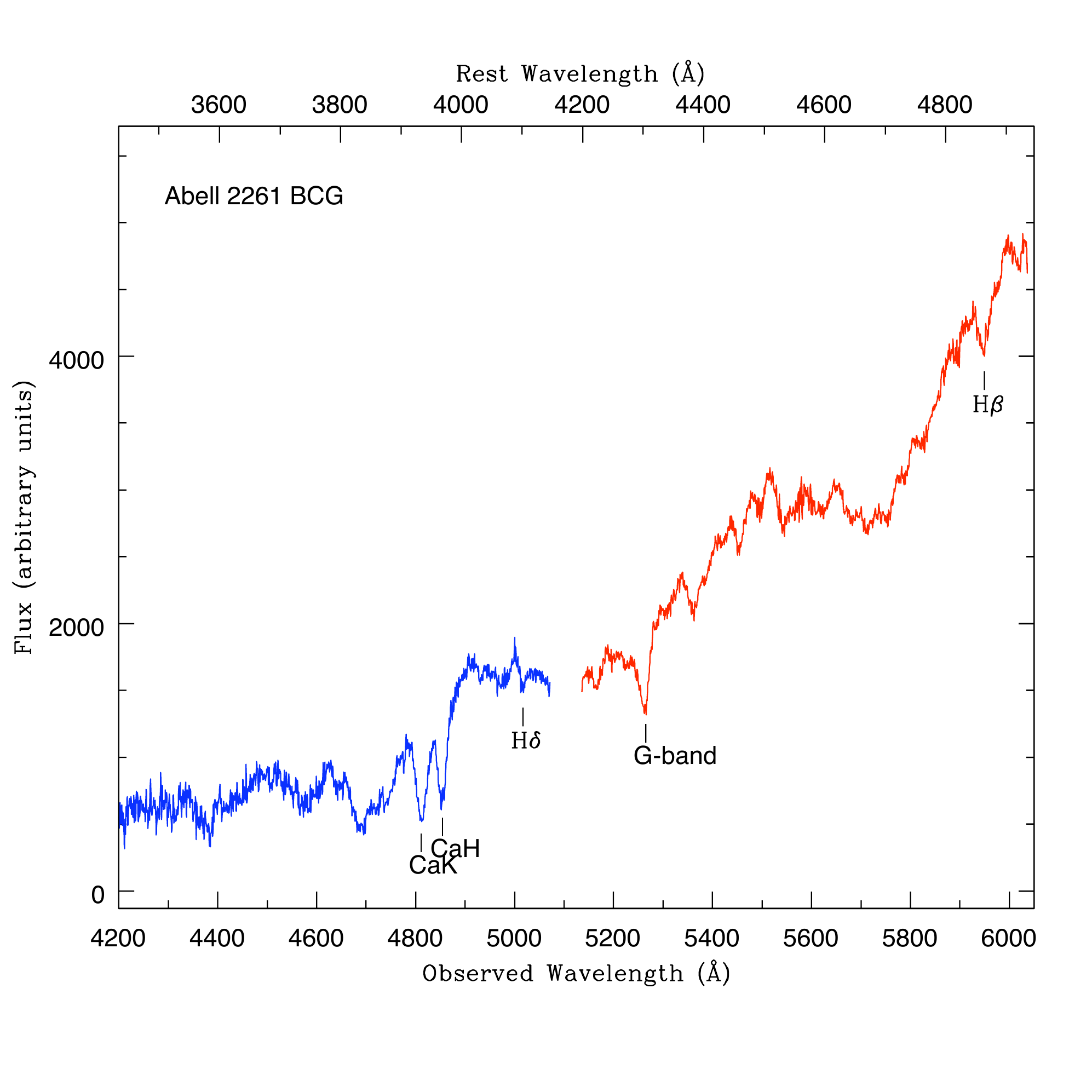}
\caption{Spectrum of the BCG in Abell 2261. Its redshift is $z = 0.2232$. 
This spectrum is from the GMOS spectrograph on the Gemini-N 8-meter telescope (P.I. H. Hoekstra).
Prominent absorption features are noted. The data from the GMOS blue and red channels are denoted, respectively, by the blue and red
traces. The break at 5100\AA\ is due to a small data gap between the two channels. This spectrum was used to derive the stellar velocity dispersion of the BCG reported in this work, $\sigma = 387 \pm 16$ km s$^{-1}$.}
\label{fig:spectra}
\end{figure}

\subsection{Photometry}

The details of the image reduction and co-addition of the Abell 2261
{\it HST} observations are given in \citet{clash} and \citet{Coe12}.
The analyses of the BCG profile here is based on the F606W and F814W
HST CLASH images.
The ACS photometry is on the AB magnitude system but we will express
much of the reduced photometry in
rest-frame V-band (Vega-based) for compatibility with the \citet{l07a}
analysis of central galaxy photometry.
We use a suite of \citet[hereafter BC03]{bruz}
synthetic stellar population models to derive linear photometric 
transformations from the observed,
extinction-corrected ACS/WFC magnitudes and colors to the rest-frame Johnson V-band, $V_{\it o}$.
The BC03 models used to empirically derive the transformation
equations have the following parameters: 
exponential star formation rate e-folding times,
$\tau$, of 0 Gyr (SSP model), 0.2 Gyr, 0.6 Gyr, and 1.0 Gyr; 
metallicities of 0.25$Z_{\odot}$, 1.0$Z_{\odot}$, and 2.5$Z_{\odot}$,
and ages from 2 Gyr to the age of the universe at
the cluster redshift in 0.5 Gyr intervals. These models more than span the range of
observed cluster galaxy SEDs.
To compute photometric transformations,
we take the BC03 spectral energy distributions (SEDs) for
each of the above models and compute the observed
photometry and colors at the cluster redshift and then use the same SEDs
to compute the rest-frame Johnson V magnitude. 
The transformation equation parameters are then derived using
a linear least-squares fitting procedure. 
In addition, we compute an evolution correction, $c_{ev}$,
from a BC03 $\tau=0.6$ Gyr
solar metallicity model with a formation epoch of $z = 4.5$. 
The transformation equation is:
$$
V_o = ({\rm F814W} + c_{ev,F814W}) + 0.5186 \times
({\rm F606W - }$$
$$\ {\rm F814W} + c_{ev,F606W} - c_{ev,F814W}) + 0.1764$$
\begin{equation}
 = ({\rm F814W}) + 0.5186  \times ({\rm F606W - F814W}) + 0.4213
\end{equation}
where F606W and F814W are the Galactic extinction-corrected ACS/WFC
measurements in AB-magnitudes, $c_{ev,F606W} = 0.256$,
$c_{ev,F814W} = 0.233$, and $V_o$ is on the Vega system.
The extinction corrections used are $0.127$ mag and $0.079$ mag for the
F606W and F814W filters, respectively. The rms scatter about the transformation
equation shown in equation 1 is 0.006 mag.
We tested the robustness of the method by computing similar
transformations for F625W and (F625W-F814W) and for
F850LP and (F814W-F850LP).
They yield the same rest-frame $V_o$ values to within $\pm 0.01$ magnitudes.

The total luminosity of A2261-BCG was determined by fitting its surface photometry with an $r^{1/4}$ law.  
This is consistent with the methodology in \citet{l07a}, which will be used to provide the context for the present galaxy.
The BCG is highly luminous with a total absolute magnitude of $-24.70$
in the V-band (see Table \ref{tab:GalPhot}).
Because we want to compare this to the $z=0$ sample of BCGs,
this luminosity includes
an evolutionary correction of $+0.26$ mag to account
for the aging of the stellar population since the $z=0.22$ epoch.
Even then, A2261-BCG is among the most luminous BCGs known \citep{pl}. 

Four sources fall within the outskirts of the core.
Their coordinates and photometric properties are summarized in
Table~\ref{tab:GalPhot}. 
The brightest source, ``knot 3," is well resolved. Its profile was measured
simultaneously with the BCG (as is described in $\S\ref{sec:sphot}$)
and has a roughly exponential form.
The close pair of compact sources north-east of the core center,
knots 1 and 2,
are both marginally resolved with $~0\asec05$ half-intensity radii.
The faint source south of the core, knot 4, is unresolved.
The photometric errors
in Table~\ref{tab:GalPhot}  given here include both
the random and estimated systematic errors. The random errors (photon shot noise,
read noise, dark current) are in the range 0.005 mag to 0.010 mag. The systematic errors
are relatively small but still about 2 to 4 times
 larger than the corresponding random errors. This is due to 
spatial variations in the background
level after the BCG model is subtracted from the image. To estimate the amplitude of the systematic
error we compared the knot photometry derived from the BCG-subtracted image with that 
derived by using a local sky subtraction (instead of BCG subtraction) where the background is 
estimated from an annulus with an inner radius of $0\asec6$ and an
outer radius of $1\asec2$ centered on each knot. The two measures of the photometry typically agree to
within 0.015 mag to 0.020 mag. Based on this, we adopt 0.018 mag as an estimate of the systematic error.

\begin{deluxetable*}{ccccccc}
\tabletypesize{\scriptsize}
\tablewidth{0pt}
\tablecaption{\label{tab:GalPhot}Abell 2261 BCG and Bright Knot Astrometry and Photometry}
\tablehead{
\colhead{Object}&
\colhead{R.A.}&
\colhead{Dec.}&
\colhead{F814W} &
\colhead{F606W-F814W}&
\colhead{V Mag}&
\colhead{M$_V$}\\
\colhead{ID}&
\colhead{(J2000)}&
\colhead{(J2000)}&
\colhead{(AB mag)$^a$}&
\colhead{(AB mag)$^a$}&
\colhead{(Vega mag)$^b$}&
\colhead{(Vega mag)$^c$}
}
\startdata
BCG    & 17:22:27.18 & $+$32:07:57.30 & 14.69        & 0.82  &  15.53 & $-24.70$ \\
             &                     &                           & ($<0.01$) & ($<0.01$) & (0.01) & (0.02) \\
Knot 1 & 17:22:27.23 & $+$32:07:57.65 &  21.53 & 0.86  &  22.15 & $-17.85$\\
             &                     &                           & (0.02) & (0.04) &  (0.06) & (0.06)   \\
Knot 2 & 17:22:27.21 & $+$32:07:57.56 &  21.89 & 0.85  &  22.50 & $-17.49$\\
             &                     &                           & (0.02) & (0.04) & (0.06)  &  (0.06)  \\
Knot 3 & 17:22:27.13 & $+$32:07:57.59 &  20.20 & 0.83  &  20.80 & $-19.20$\\
             &                     &                           & (0.02) & (0.04) & (0.06)  &  (0.06)  \\
Knot 4 & 17:22:27.14 & $+$32:07:55.85 &  22.93 & 0.97  &  23.61  & $-16.63$\\
             &                     &                           & (0.02) & (0.04) &  (0.06) &   (0.06) \\
\vspace{-0.1in}
\enddata
\tablenotetext{a}{Corrected only for Galactic Extinction.}
\tablenotetext{b}{Corrected for Galactic extinction and K-dimming (assuming sources are at the cluster redshift).}
\tablenotetext{c}{Corrected for Galactic extinction, K-dimming, and passive 
evolution from $z=0.224$ to $z=0$. See text for details}
\end{deluxetable*}

\subsection{Spectroscopy and The Central BCG Velocity Dispersion}

Moderate resolution spectroscopy of
the BCG exists from the Sloan Digital Sky Survey (SDSS) DR8 and
from the Gemini Observatory data archive.
The Gemini data are deep long-slit spectra with GMOS-N from
the program GN-2008A-Q-103 (P.I. Chris Bildfell).
The Abell 2261 observations were obtained on March 16, 2008.
The GMOS-N long-slit, with $0\asec75$ width, was placed on the BCG and oriented at a position angle
that intersects two of the three bright knots in the northern part
of the core (but not the brightest one).
The spectra are split into dual red and blue channels to optimize sensitivity.
The full spectral range is $\sim 4140 - 6040$\AA\
with a spectral resolution of $R \sim 1000$ at the blue end.
The 1D co-added GMOS spectrum is shown in Figure \ref{fig:spectra}.

We measured the stellar velocity dispersion ($\sigma$) within the
core of A2261-BCG from the GMOS spectrum using the IDL code
{\tt pPXF} developed by \citet{vdisp},
which fits a linear combination of template spectra to the observed
galaxy spectrum to minimize template mismatch.  The template spectra
are provided by the \citet{vaz} single
burst models for a range of ages and metallicities, which are based on
the empirical MILES stellar library \citep{sanchez}.

The fits are computed separately in the red and blue channel spectra,
covering rest-frame wavelengths 3700--4150\AA\ and 4250--4700\AA,
respectively in order to avoid regions near the chip edges where the
wavelength solution is less reliable.  The bright skyline at the
5682--5688\AA\ Na doublet is masked in the fitting procedure, and the
fits are computed weighting all pixels equally.  We find consistent
values of $\sigma = 393 \pm 13$ km s$^{-1}$ on the blue side and
$\sigma = 380 \pm 8$ km s$^{-1}$ on the red side.
The higher-order non-Gaussian velocity moments $H3$
and $H4$ derived from the fits are negligible.
We therefore adopt $\sigma = 387 \pm 16$ km s$^{-1}$ as an estimate of the
ensemble stellar velocity dispersion, including systematic errors.
The Gemini velocity dispersion value 
is in excellent agreement with the $388 \pm 19$ km s$^{-1}$
velocity dispersion estimate from the DR8 SDSS database.
This is among the highest central velocity dispersion values known.

\begin{figure*}
\plotone{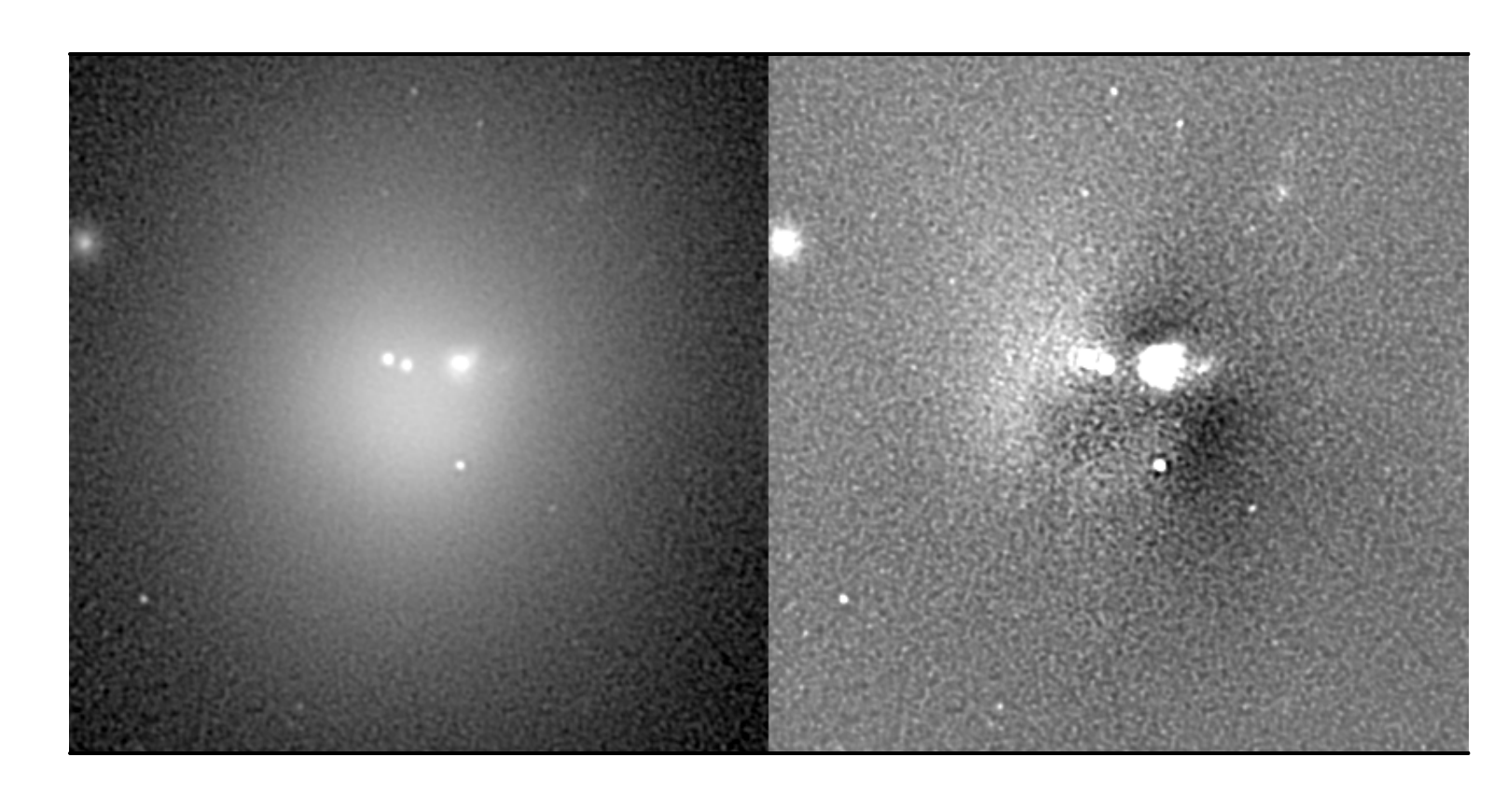}
\caption{The left panel shows the center of the F814W image of the BCG after
\citet{lucy}-\citet{rich} deconvolution.
The region shown is $12''\times12'';$ ($43.2\times43.2~$ kpc)
the intensity scale is logarithmic. North is at the top and east to the left.
The right panel shows the residuals after subtraction of a model
reconstructed from the surface photometry of the BCG.  The over all
structure of the residuals is a dipole pattern of positive residuals
NE of the core and negative residuals to the SW.  This suggests that
the core is slightly displaced from the surrounding envelope in the
SW direction.}
\label{fig:core}
\end{figure*}

\section{The Central Structure of A2261-BCG}

\subsection{The Morphology of the Core \label{sec:sphot}}

The surface brightness profile of A2261-BCG was measured from the
F814W image deconvolved with
20 cycles of \citet{lucy} - \citet{rich} deconvolution to correct
for the blurring of the PSF.
The deconvolved image of the core is given in Figure \ref{fig:core}.
Deconvolution works well on {\it HST} images for recovering estimates
of the intrinsic light distributions of galaxies \citep{l98, l05}.
Given the angularly-large and flat-profile of the A2261-BCG core, the effects
of deconvolution in the present context are extremely modest.
The increase in the central surface brightness within the core
after deconvolution is only $\sim3\%,$
with similar effect on the measured angular size of the core.
The real import of deconvolution for A2261-BCG is to recover the
forms of the compact sources falling around the core,
and to reduce the effects of their scattering ``wings'' on the core, itself.

Figure \ref{fig:core} also shows the residuals obtained after a 2D model
reconstructed from the brightness profile of the core was subtracted
from the image.
The profile for $r>0\asec5$ was measured using the \citet{snuc} algorithm
(operating in the {\tt xvista} image processing system),
which solves for the overlapping light distributions of multiple galaxies;
the high-resolution algorithm of \citet{l85} was used interior to this.
The surface brightness profile is presented in Figure \ref{fig:sb}.
The profile ratifies the visual impression that the core is essentially flat.
In contrast to the typical central structure of
most elliptical galaxies (but not all --- see \citealt{l02}),
there is no sign of any rising cusp in starlight as $r\rightarrow0.$
The lack of a cusp actually made it difficult to determine the
precise center of the galaxy.
We did this by taking an intensity centroid over the entire core.

At radii well outside the core, {\it HST}-based surface brightness
profiles become vulnerable to systematic sky measurement
errors and so on, due to the low intensity levels of the outer isophotes, the
angularly-small pixels, and limited fields of the cameras.  To better
characterize the envelope of A2261-BCG, we augmented the ACS
photometry with surface photometry measured from a ground-based R-band
image obtained with the Suprime-cam imager on the Subaru 8-m telescope.
The Subaru-derived
profile was scaled to match the ACS profile over $5''<r<8'',$
then blended with it over the same interval, and used solely as the
brightness distribution for $r>8''.$  The Subaru profile extends
to $\sim 24''$ and equivalent V-band surface brightnesses
fainter than 25 mag/${\rm arcsec^2}.$

\begin{figure}
\plotone{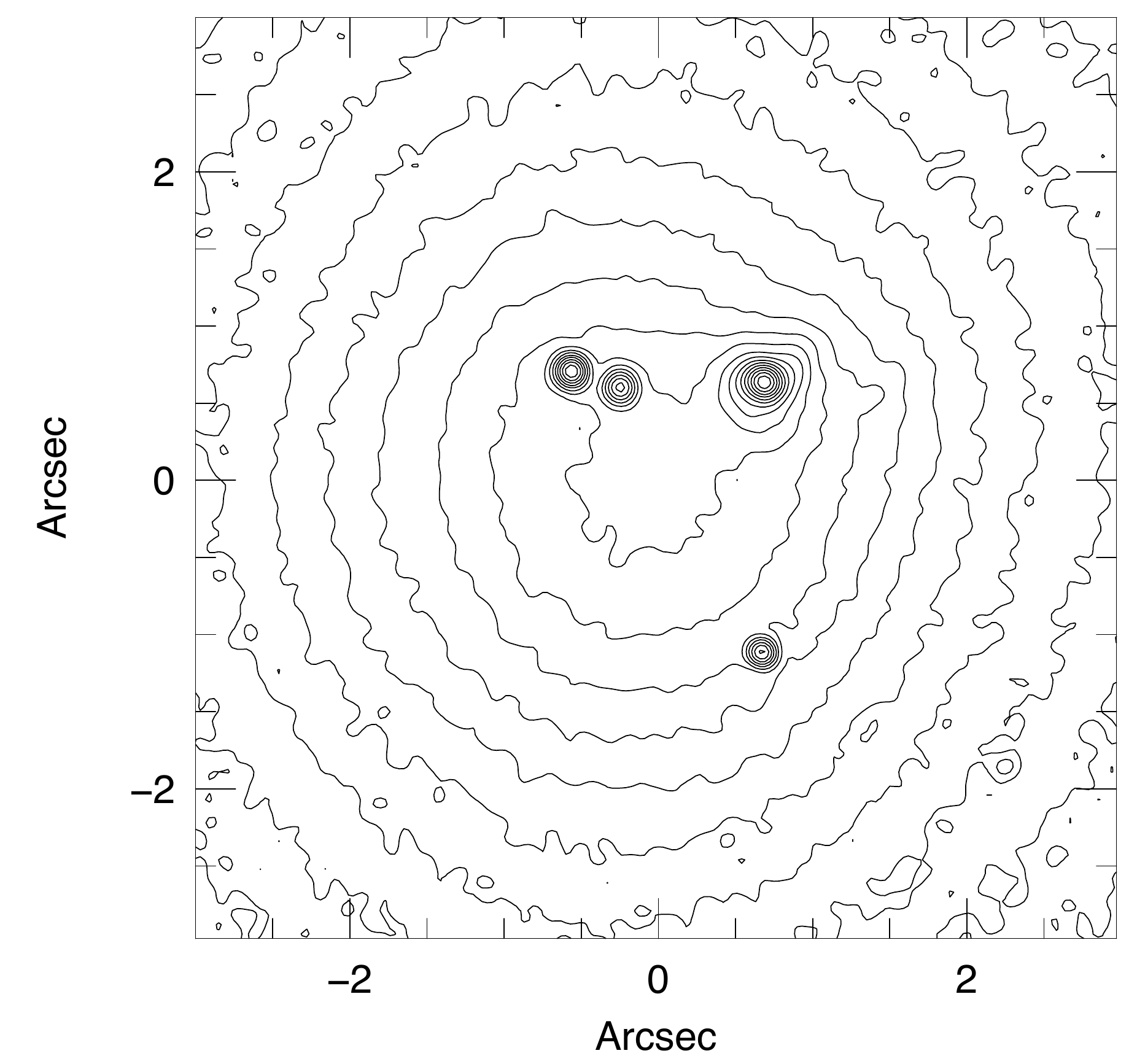}
\caption{A contour plot of the core of A2261-BCG.  The contour
levels have an arbitrary zeropoint, but are spaced by 0.25 mag in surface
brightness. North is to the top and east to the left.
Note that the contour levels are closer together in the SW
direction outside the core than they are in the NE, supporting the
dipole-like residual pattern seen in Figure \ref{fig:core} and the
conclusion that the core is displaced to the SW relative to the envelope
center.}
\label{fig:con}
\end{figure}

Despite the smooth structure of the profile, the residuals within the core
show an number of interesting features.
In addition to the presence of the four sources noted earlier,
the residuals show a dipole-like
pattern outside the central flat portion of the profile.
The simplest interpretation is that the core is slightly displaced from
the center of the surrounding envelope by $\sim0\asec2$ (0.7 kpc) towards
position angle $\sim300^\circ.$  This conclusion is supported by the
contour map of the core, shown in Figure \ref{fig:con}.  The
contour lines outside of the core to the SW are clearly spaced
closer together than those in the NE direction.

\subsection{Analysis of the Surface Brightness Profile}

We describe the profile with a ``Nuker-law'' \citep{l95},
\begin{equation}
I(r)=2^{(\beta-\gamma)/\alpha}I_b\left({r_b\over r}\right)^{\gamma}
\left[1+\left({r\over r_b}\right)^\alpha\right]^{(\gamma-\beta)/\alpha},
\label{eqn:nuker}
\end{equation}
which models the surface-brightness
distribution as a ``broken'' power-law.
The envelope profile has the form $I(r)\sim r^{-\beta}$ as
$r\rightarrow\infty,$ while the inner cusp has $I(r)\sim r^{-\gamma}$
as $r\rightarrow0,$ with the transition radial-scale provided
by the ``break-radius,'' $r_b.$  The ``speed'' of transition
between the envelope and inner cusp
is provided by $\alpha,$ while $I_b,$ the surface-brightness
at $r_b$ gives the overall intensity normalization.

\begin{figure}
\plotone{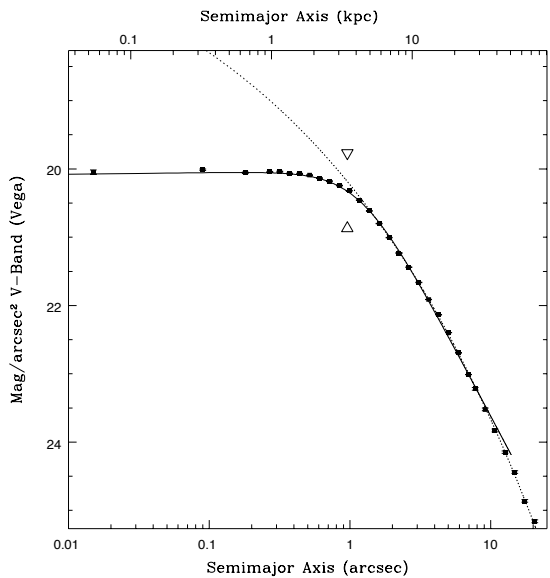}
\caption{The central surface brightness profile of A2261-BCG
as measured (solid points) is shown with two ``Nuker-law'' profile
fits \citep{l95}. The error bars are smaller than the points,
but for the central few measurements.
For comparison to previous studies the profile is normalized to
$z=0$ V-band (Vega).
The solid line is the best-fitting Nuker-law
and features a slightly depressed ($\gamma=-0.01$) cusp as $r\rightarrow0.$
The dotted line is an
is an $r^{1/4}$-law (an $n=4$ S\' ersic-law) fitted to the envelope.
The triangles indicate the cusp-radius.}
\label{fig:sb}
\end{figure}

\begin{figure}
\plotone{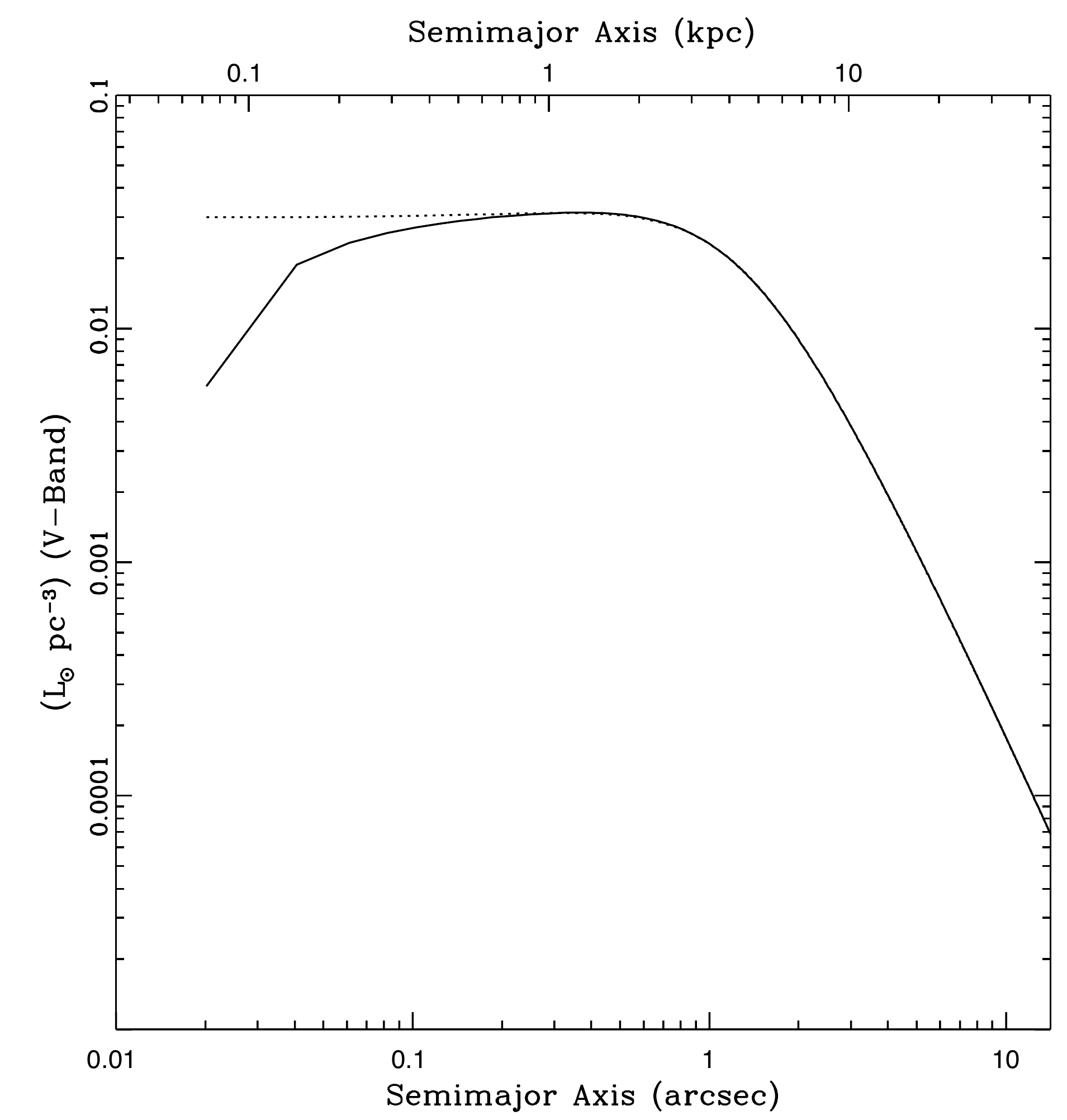}
\caption{The implied luminosity density profile of A2261-BCG.
The density profile resulting from Abel inversion of the
best-fitting Nuker-law, which features a slightly depressed ($\gamma=-0.01$)
cusp as $r\rightarrow0,$ is shown as a solid line.  The density
profile plotted as a dotted-line is based on inversion
of a Nuker-law fitted with $\gamma=0.$  The small difference
between the two profiles is greatly amplified in density as $r\rightarrow0.$
Even the density profile with $\gamma=0$ is formally very slighted
depressed as $r\rightarrow0.$ The difference between the two models
is not significant.}
\label{fig:den}
\end{figure}

The best-fitting Nuker-law is plotted in Figure \ref{fig:sb},
and has the parameters $\gamma=-0.01,$
$\beta=1.56,$ $\alpha=2.41\pm0.18,$
$r_b=1\asec20,$ and $I_b$ of $19.72$ in F814W(AB).
The fit was conducted for $r<8'',$ beyond which the the envelope
falls away slightly faster than a pure power-law with radius.
As can be seen, the fit over this domain is excellent, with an rms
residual of 0.02 mag.
The parameters are not strongly dependent on
the domain of the profile fitted, in any case.  For a fit
conducted limited to $r<3\asec7,$ or half of the nominal domain,
$r_b=1\asec15$ is recovered, only a 4\%\ decrease from the nominal value.
The break radius, $r_b,$
corresponds to the point of maximum logarithmic curvature,
and thus can also be estimated by non-parametric methods.
\citet{l07a} did this for a the large sample of core galaxies
presented in \citet{l05}, finding the Nuker-fit and non-parametric
measurements to agree well over a large range in angular core-size,
with no biases.

The negative $\gamma$ actually implies that the
surface brightness decreases slightly as $r\rightarrow0.$
We also performed a Nuker-fit forcing $\gamma=0,$
which is essentially identical to the nominal fit, but for the central point.
In this case, we recover $\beta=1.55,$ $\alpha=2.50,$
$r_b=1\asec30,$ and $I_b$ of $19.72.$
The $\gamma=0$ model falls right on the upper error-bar
of the central point and thus coincidentally serves as a $1\sigma$
confidence bound on the central profile.

The center of A2261-BCG may have a luminosity {\it density} profile
that actually decreases as $r\rightarrow0.$
This occurs in a number of galaxies,
including a few BCGs \citep{l02, l05}.
An Abel inversion of the Nuker-model for A2261-BCG
with $\gamma=-0.01$ indeed shows that $r=0$ corresponds
to a local minimum in stellar density (Figure \ref{fig:den}).
Even inversion of the $\gamma=0$ model of the brightness profile
formally implies a $\sim4\%$ decrease in luminosity density
at the {\it HST} resolution limit.  This can be understood
as a consequence of the lack of a cusp in combination with a
sharply-defined core.  With a generic surface brightness profile
of the form, $I(r)\propto\left(1+r^\alpha\right)^{-\beta/\alpha}$
(both $\alpha$ and $\beta$ positive),
the derivative of the profile has the form $I(r)'\propto r^\alpha$
for $r<<1.$  When used with the standard Abel-transform,
this implies a formal density of zero at $r=0,$ when $\alpha>2,$
a condition satisfied for the A2261-BCG brightness profile.

Given, however, the modest central decrease in density implied
by even the $\gamma=-0.01$ model, which just occurs near the resolution limit,
it is not possible to say with confidence
if A2261-BCG is yet another example of a ``hollow core.''
At the same time there is no sign
of any centrally {\it rising} cusp in central surface brightness,
and an even {\it constant} density core over nearly two decades
in radius is unusual \citep[see the collection of core density
profiles in][]{l07b}.

\subsection{The A2261-BCG Core Compared to Those of Other Early-Type
Galaxies}

\begin{figure*}
\plotone{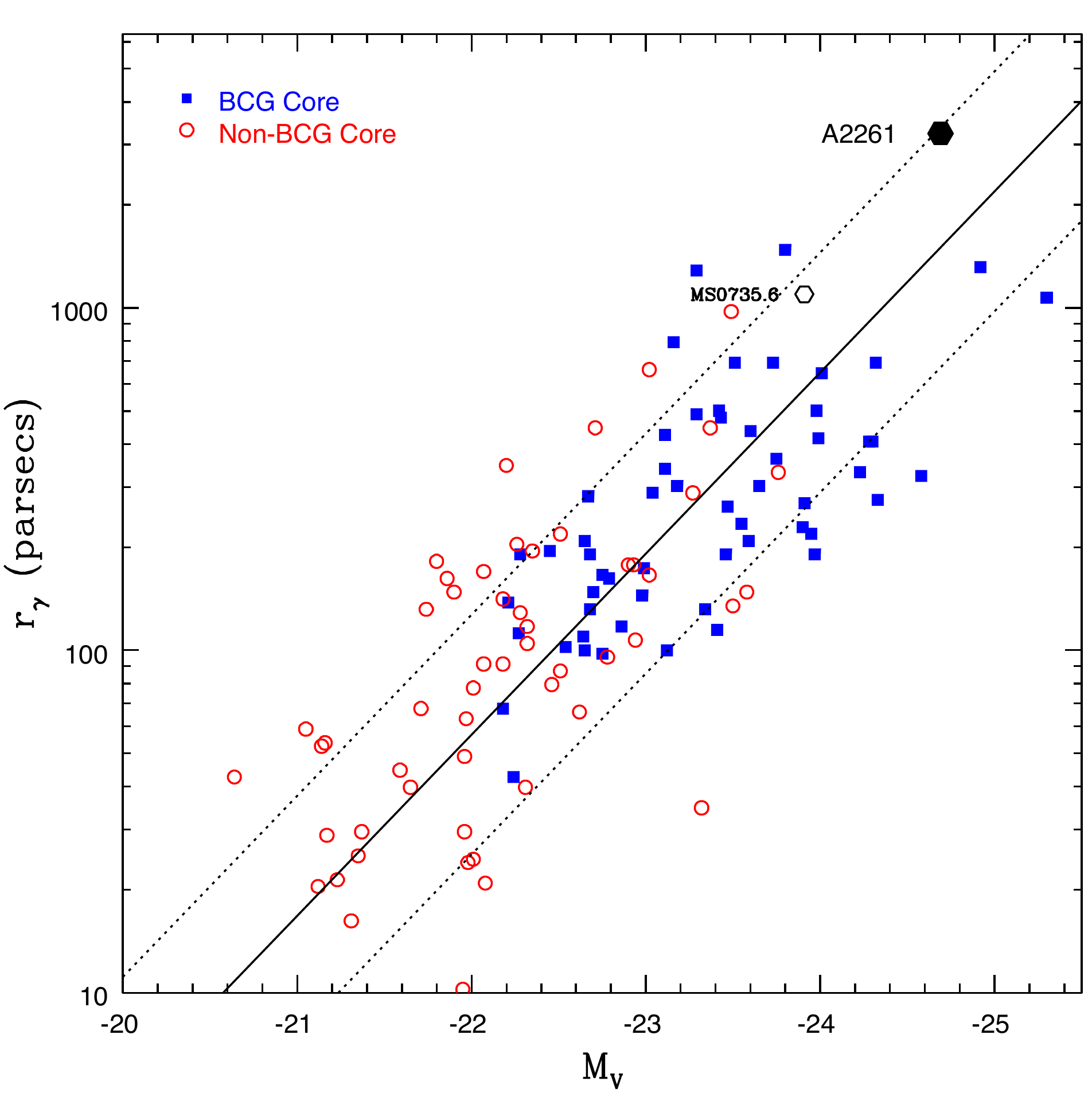}
\caption{The relationship between cusp-radius
and total galaxy luminosity (V-band Vega-based).
The galaxy sample plotted was assembled in \citet{l07a} from
a variety of sources (the figure is adopted from Figure 5 in
that paper).  The BCGs in particular come from the \citet{laine} sample.
The \citet{l07a} $r_\gamma-L$ relationship (also given in equation
\ref{eqn:r_mv}) is plotted; the dotted lines indicate $\pm1\sigma$
scatter about the mean relationship.
A2261-BCG is plotted at the top,
clearly has a cusp-radius larger than all other galaxies in the sample.
The large core in the MS0735.6+7421 BCG discovered by \citet{mac}
is also plotted for comparison.}
\label{fig:lrc}
\end{figure*}

Figure \ref{fig:lrc} shows that the core of A2261-BCG is the largest
core yet seen among extensive surveys of local galaxies.
The figure is an adaptation of a figure in \citet{l07a}, which
shows the relationship between the core ``cusp-radius,'' $r_\gamma,$
and total V-band luminosity, $L_V$ for a composite sample
of {\it HST} studies of the central structure of early-type galaxies
\citep{l95, f97, quil, rav, rest, laine, l05}.\footnote{The V-band
used in \citet{l07a} and cited in the remainder of the paper is Vega-based.}
The \citet{laine} study, in particular,
focussed on the central structure of BCGs, and is thus particularly
useful for placing the core of A2261-BCG in context.

The cusp-radius is the angular or physical scale at which the
local logarithmic slope of the surface brightness profile reaches
a value of $-1/2,$ as the profile transitions from the steep envelope
to the shallower inner-cusp.  In terms of the Nuker-law parameters,
\begin{equation}
r_\gamma\equiv r_b\left({1/2-\gamma\over\beta-1/2}\right)^{1/\alpha}.
\end{equation}
This scale was first introduced by \citet{carollo} and adopted
in the analysis of \citet{l07a}, who demonstrated that it
gave tighter relationships between the core and global properties
of the galaxies, than the direct use of the break-radius, $r_b,$ did.

For A2261-BCG, the $\gamma=-0.01$ Nuker-law fitted yields
$r_\gamma=0\asec89\pm0\asec02,$
or $3.2\pm0.1$ kpc at the distance of A2261-BCG
(the $\gamma=0$ fit also gives $r_\gamma=0\asec89$).
As a check, this value is consistent with a simple estimate
of $r_\gamma=0\asec90,$ made by fitting a parabola to the core
over $0\asec7<r<1\asec4.$ The corresponding
``cusp-brightness,'' $I_\gamma,$ the surface brightness at $r_\gamma,$
is 19.52 mag/arcsec$^2$ (F814W) or 20.29 in rest-frame $V$-band
corrected for interstellar extinction.
As shown in Figure \ref{fig:lrc},
this $r_\gamma$ is over twice as large as the largest BCG cores seen in the
\citet{laine} sample.  It is also three times larger than the
large core in MS0735.6+7421-BCG noted by \citet{mac}.
The A2261-BCG core falls on the high-side of the \citet{l07a}
relation between core-size and total galaxy-luminosity,
\begin{equation}
\log (r_\gamma/{\rm pc})=(1.32\pm0.11)(-M_V-23)/2.5+2.28\pm0.04.
\label{eqn:r_mv}
\end{equation}
The scatter in $\log (r_\gamma)$ about this relation is 0.35 dex;
A2261-BCG falls 0.33 dex above the relation, corresponding to
a $1.0\sigma$ deviation.

\section{Conjectures About Supermassive Black Holes in the Core}

\subsection{Estimates of the Black Hole Mass From the Core Size}

If cores are created by the scouring action of a binary black hole,
and the core size is indicative of the mass of the binary,
then the implied black hole mass for A2261-BCG must be
extremely large. \citet{mcc} recently identified
two black holes with masses $10^{10}M_\odot$ or greater in the
BCGs A1367-BCG=NGC 3842 and A1656-BCG=NGC 4889, which have cusp-radii
of only $0.3$ kpc and $0.7$ kpc, respectively \citep{l07a}, implying that
$M_\bullet$ in A2261-BCG is yet larger 
than that found in those two galaxies.

The $M_\bullet-L$ and $M_\bullet-\sigma$ relations between
nuclear black hole mass and galaxy luminosity \citep{d89,kr95,mag}
or velocity dispersion \citep{fm,g00} give some guidance
on what to expect for the central black hole mass in
A2261-BCG, completely apart from its core structure.
The $M_\bullet-L$ and $M_\bullet-\sigma$ (elliptical galaxy
only sample) relations from the comprehensive analysis of \citet{g09} predict
$M_\bullet=6\times10^9M_\odot$ and $2\times10^9M_\odot,$ respectively.
The analogous \citet{mcc} $M_\bullet-L$ and $M_\bullet-\sigma$ relations,
which as noted above include BCG black hole masses,
predict somewhat higher values of $1.1\times10^{10}M_\odot$
and $5\times10^9M_\odot.$  It should be noted that
A2261-BCG is more luminous and
has a higher velocity dispersion than any galaxy that actually
has a black hole mass measurement, thus the predictions
are extrapolations. 

A different approach is to use the fundamental-plane of black hole activity
\citep{merlot}, which derives $M_\bullet$ from a combination
of 5 Ghz core radio luminosity and $2-10$ keV nuclear X-ray emission.
\citet{hl12} use this methodology to estimate
$M_\bullet=2.0^{+8.0}_{-1.6}\times10^{10}M_\odot$ for A2261-BCG.

Unfortunately, estimates of black hole mass implied by the
scale of the core, itself, in A2261-BCG will be even more uncertain
extrapolations.  \citet{l07a} presented a number of relationships
between $r_\gamma$ and $M_\bullet,$ based on how the sample
of core galaxies with real $M_\bullet$ determinations was analyzed.
If $r_\gamma$ and $M_\bullet$ are fitted symmetrically, then
\citet{l07a} find $r_\gamma\propto M_\bullet^{0.8},$ which
predicts $M_\bullet\sim4\times10^{10}M_\odot$ for A2261-BCG.
If on the other hand $M_\bullet$ is treated as the independent
variable, then \citet{l07a} find $r_\gamma\propto M_\bullet^{1.5},$
which predicts a more modest $M_\bullet\sim7\times10^9M_\odot.$
The large difference between the two estimates is due to the small
number of systems contributing to the \citet{l07a} $r_\gamma-M_\bullet$
relationships and the large scatter between the two parameters.

An alternate approach advocated in several papers
is to relate estimates of the ``mass-deficit,'' $M_d,$ that is the mass
in stars that was ejected to generate the core, to $M_\bullet.$
Unfortunately, this methodology at present also leads to large
uncertainties in the estimated $M_\bullet$ for A2261-BCG.
\citet{kb} derive a relationship between $L_d$ and $M_\bullet$ for
galaxies of the form $L_d\propto M_\bullet,$ where $L_d$ is the
observed {\it starlight} rather than inferred mass deficit, and
is measured by integrating the central difference between the
observed surface brightness profile and an inward extrapolation
of a S\' ersic-law fitted to the envelope of the galaxy.

The best fit of a S\' ersic law to the envelope of A2261-BCG has
$n=4.1,$ or essentially the classic $r^{1/4}$-law form.
An $r^{1/4}$-law fitted to the envelope gives $M_V\approx-20.8,$
implying $M_\bullet\sim1.1\times10^{10}M_\odot,$
using the \citet{kb} relation.
This later $L_d$ is identical to the $M_V\approx-20.8$ derived
by the simple estimate of ``core-luminosity,'' $L_\gamma\equiv\pi I_\gamma
r_\gamma^2$ used by \citet{l07a} as a proxy for $L_d$ that avoids
the need for the careful selection and evaluation of a ``pre-scouring''
reference surface brightness profile.

\subsection{Ejection of the Central Black Hole}

One intriguing possibility is that the large flat core has
resulted from the central ejection of its nuclear black hole.
In this scenario, the galaxy most likely had a large core to begin
with, which would have been enlarged by the ejection of the central
black hole.  The large core would have been generated by
a smaller black hole than that directly implied by
the presumption that scouring by the binary black hole did all the work.

\citet{dm04} and \citet{bk04} show that
a core can be generated directly in a ``power-law'' galaxy
(a system that initially has a steep central cusp) when the components
of a binary black hole merge and the remnant is ejected
by the asymmetric emission of gravitational radiation.
\citet{gm08} simulated this scenario, finding that
that ejection can substantially enlarge a pre-existing core,
leading to the inference of an exceptionally large mass deficit.
Given the high luminosity of A2261-BCG, it is extremely likely
that it would have a large core prior to any
ejection of a central black hole.
\citet{gm08} also show that the central stellar density profiles
can become extremely flat after ejection of the black hole,
similar to what is seen in Figure \ref{fig:den}.

The physical displacement of the
core from the envelope center, noted above, is indicative
of a local dynamical disturbance.
This is likely to be a relatively recent event.
An ejected black hole would be trailed by a strong dynamical-friction
wake as it leaves the core.  In effect, it would ``pluck'' the core,
which would then oscillate for a few crossing times,
$t_c\sim r_\gamma/\sigma$ or $\sim10^7$ yr for A2261-BCG.
At the same time, this need not mean that the ejection of the
black hole, itself, needs to be as recent.  If the ejected black
hole remains on a radial orbit that periodically returns to the core,
then the disturbance might only be due to a recent passage
(Merritt, private communication).

Real proof of the ejected black hole hypothesis would be to find
direct evidence of the ejected black hole, itself.
\citet{dm09} show that the ejected black hole carries along with it a ``cloak''
of stars that had previously been closely bound to it.
The ejected black hole and associated stars would
somewhat resemble a globular cluster or ultra-compact dwarf galaxy.
The key diagnostic for such a system
would be the extremely high velocity dispersion
of the tightly bound stars.
An obvious question then for the
present case is if any of the four sources proximal to
the core might be such an object.

An important point is that the stellar ``cloak'' is most
likely to be considerably less massive than the black hole.
For a black hole ejected from a core, \citet{dm09} find
\begin{equation}
{M_b\over M_\bullet}\approx2\times10^{-2}\left({\sigma\over200~\rm{km~s^{-1}}}
\right)^{5/2}\left({V_k\over10^3~\rm{km~s^{-1}}}\right)^{-5/2},
\end{equation}
where, $M_b$ is the mass of stars bound to the black hole,
$\sigma$ is the velocity dispersion of the galaxy,
and $V_k$ is the ``kick'' velocity with which the newly merged
binary black hole is ejected from the center of the galaxy.
In the case of A2261-BCG with $\sigma=387\  \rm{km~s^{-1}},$
the coefficient in the above equation becomes 0.1.
While we have echoed the nominal $V_k=1000~{\rm km~s^{-1}}$ from
\citet{dm09}, the real value depends on the unknown mass ratio and spins
of the two black holes within the merging binary and can range from
essentially zero to a few thousand km~s$^{-1}$.
The only way to constrain $V_k$ better would be
to identify the remnant and measure its line-of-sight velocity.

\citet{dm09} also show that
the effective radius of the cloak is also related to the kick velocity.
For a black hole initially in a core,
\begin{equation}
R_e\approx43\left({M_\bullet\over10^{10}M_\odot}\right)
\left({V_k\over10^3~\rm{km~s^{-1}}}\right)^{-2}~\rm{pc}.
\end{equation}
Given the expected scale and mass of the stellar cloak,
the best candidate for an ejected
black hole would be the least luminous of the four sources,
the unresolved point source south of the core.  The most luminous
of the four sources has an implied luminosity of $L_V\sim4\times10^9
L_\odot,$ and a substantially larger physical extent than the number given
above.  When converted to a likely mass, the source carries
a substantial fraction of the putative $10^{10}M_\odot$ black hole,
thus not matching the expectation that $M_b<<M_\bullet.$
Morphologically, the source also resembles any number of other
small galaxies visible within the neighborhood of the BCG.
The two less luminous paired-sources, also north of the core,
are more consistent with a possible cloak, but even their
compact sizes are large for the expectations enumerated above.

Of course if $V_k$ is large, and the ejection happened long
ago, then the remnant would be unlikely to be within the core,
and the candidate list would have to be extended to sources
at much larger distances from the center of the BCG.
The time scale for exiting the core is simply $r_\gamma/V_k,$
which is likely to be substantially shorter than $t_c.$
Conversely, a small kick favors larger and more extended cloaks,
and increases the likelihood that the remnant would remain
close to the core in the event that the kick was insufficient
to unbind the black hole from the BCG, itself.
If the offset core is indeed due to an ejected black hole,
then the short lifetime of the offset would imply that the
remnant should be relatively close by.

\section{A2261-BCG and the Formation of Cores}

There is a rich tradition in observational astrophysics
of using the extreme member of an ensemble to understand the
origin of the ensemble over all.  The role of A2261-BCG
as a witness to the mechanisms that form cores hinges
on whether its core is ``normal'' or not.  A2261-BCG bests
its closest rival, NGC 6166, for having the largest core by a factor of two.
More generally, the core of A2261-BCG stands out from all galaxies in
the subset of 57 BCGs in the \citet{l07a} composite sample of
early type galaxies imaged with {\it HST} (the great
majority of the BCGs were provided by \citealt{laine}).
Still, the core $r_\gamma=3.2$ kpc does not fall far above the
\citet{l07a} $r_\gamma-L$ relationship, given that A2261-BCG also has an
unusually large total luminosity --- even as compared to other BCGs.
A2261-BCG might be expected to harbor a $\sim10^{10}M_\odot$ black
hole, based on its location within the $M_\bullet-L$ and
$M_\bullet-\sigma$ relations, thus the formation of core in this
galaxy would {\it a priori} be expected to a limiting case
for the formation of the core by core scouring, the standard hypothesis.

The core of A2261-BCG was not at first noted for its size,
as much as its unusual appearance, however.  Cores generally have at least
a weak central surface brightness cusp, but in A2261-BCG there is no
obvious center to the core.
Its central stellar density profile is perfectly flat,
or perhaps even depressed at the center.
While there is a radio source coincident
with the core, there is no optical AGN counterpart or central
nuclear star cluster --- nothing suggests
that A2261-BCG is hosting anything like a $\sim10^{10}M_\odot$ black hole
at its center.  If binary black holes do scour out cores,
then on occasion they must also merge and be ejected from the core,
thus causing it to rebound and expand to an even larger size
than it possessed at the completion of scouring.
An attractive feature of this scenario is that it may account
for large cores that may be difficult to explain by scouring alone.
The simulations of \citet{dm04}, \citet{bk04}, and \citet{gm08}
suggest that the core of A2261-BCG matches the
expectations of what a core that has ejected
its central black hole looks like.

There is a caveat, however.  Cores with ejected black holes may be ephemeral.
\citet{f97} note that a central black hole can act as ``guardian,"
``protecting" an existing core from being infilled by the
central cannibalism of less luminous, but centrally denser galaxies.
A2261-BCG lives in a rich environment.
Figure \ref{fig:den} shows that the central mass density
of the core is $<0.1~{\rm M_\odot~pc^{-3}},$ which is extremely
diffuse in comparison to the denser cores of less luminous galaxies.
Without a central black hole in the core of A2261-BCG,
the nuclei of galaxies merging with the BCG would readily
settle intact into its center \citep{hbr}.
The four sources projected against the core, are a reminder that the galaxy may
cannibalize its neighbors, even if the sources in question,
may themselves not be at risk for this.  Again, if A2261-BCG did
lose its central black hole, this may have been a relatively recent event.

In the end, how we use A2261-BCG to test the theory of core formation
hinges on whether or not it now hosts a supermassive black hole at its center.
The radius interior to which a black hole dominates the stellar dynamics
is $r_\bullet\equiv GM_\bullet/\sigma^2.$  For a
$\sim10^{10}M_\odot$ black hole and $\sigma=387~{\rm km ~s^{-1}},$
$r_\bullet\approx300~{\rm pc}$ or $\approx0\asec08$ for A2261-BCG.
This scale is readily accessible using adaptive optics in the near-IR
on 10m-class telescopes, with the caveat that the extremely low
surface brightness interior to the core will demand long exposures,
even with such large apertures.
The detection of a {\it central} $M_\bullet\sim10^{10}M_\odot$
would suggest that the core is simply an extreme example of
the scouring mechanism.
Conversely, a demonstration that A2261-BCG
lacks a suitably massive black hole, or the possible demonstration that
one of the sources near the core is a cloaked black hole could
establish that the large core indeed was generated by the ejection
of its central black hole.
An improved position for the central radio source might also
determine if a black hole truly lies at the center of the core,
or instead is associated with one of the knots.

\acknowledgments

We thank David Merritt for useful conversations.
Based on observations made with the NASA/ESA 
{\it Hubble Space Telescope}, obtained at the Space Telescope Science Institute,
which is operated by the Association of Universities for
Research in Astronomy, Inc. (AURA), under NASA contract NAS 5-26555. The HST
observations are associated with GO proposal \#12066.
Also based on observations obtained at the Gemini Observatory
(acquired through the Gemini Science Archive), 
which is operated by the AURA
under a cooperative agreement with the NSF on behalf of the Gemini partnership: the National 
Science Foundation (United States), the Science and Technology Facilities Council (United Kingdom),
the National Research Council (Canada), CONICYT (Chile), the Australian Research Council (Australia), 
Minist\'{e}rio da Ci\^{e}ncia, Tecnologia e Inova\c{c}\~{a}o (Brazil)
and Ministerio de Ciencia, Tecnolog\'{i}a e Innovaci\'{o}n Productiva
(Argentina). AZ is supported by contract research {\it Internationale Spitzenforschung II-1}
of the Baden-W\"urttemberg Stiftung.

\end{document}